# Left-right asymmetry in an optical nanofiber

N.I. Petrov

Lenina Street, 19-39, Istra, Moscow region, Russia

e-mail: petrovni@mail.ru

Symmetry breaking effect for left- and right-handed circularly polarized light beams propagating in a rotationally symmetric graded-index optical fiber is presented. It is shown that a left-right asymmetry manifested as an unequal transmission for opposite circular polarizations occurs due to spin-to-spin angular momentum conversion caused by tensor interaction.

*Introduction*.- Various symmetry breaking effects arise at the propagation of a polarized light in dielectric media. Optically active materials exhibit different transmission levels for left and right-hand circular polarizations (circular dichroism). Conventional optical activity is associated with intrinsically 3D-chiral molecules, and it is the property of unequal absorption of right and left hand circular polarized light. Recently, directionally asymmetric transmission of polarized light in planar chiral structures was discovered [1]. Optical activity may also arise from extrinsic chirality. Strong optical activity and circular dichroism in non-chiral planar microwave and photonic metamaterials was demonstrated in [2]. Polarization-dependent symmetry breaking effects occur also for a light propagating in optical waveguides. It was shown in [3] that spin-orbit interaction causes asymmetry effect for depolarization of a light propagating in a graded-index fiber. Spin-dependent relative shift between right- and left-hand circularly polarized light beams propagating along a helical trajectory in a graded-index fiber was shown in [4]. In [5] this effect was observed experimentally for a laser beam propagating in the glass cylinder along the helical trajectory. Recently, a phenomenon of spin-dependent splitting of the focal spot of a plasmonic focusing lens was demonstrated experimentally [6, 7]. This shift can be regarded as a manifestation of the optical Magnus effect [8] and the optical spin-Hall effect [9, 10] which arises due to a spin-orbit coupling. The effect of spin symmetry breaking via spin-orbit interaction, which occurs even in rotationally symmetric structures, was observed in plasmonic nanoapertures [11]. Similar effect of polarization-dependent transmission through subwavelength round and square apertures was demonstrated in [12].

In this Letter, the symmetry breaking effect for left- and right-handed circularly polarized light in isotropic graded-index fiber due to tensor forces is demonstrated analytically by solving the full three-component field Maxwell's equations. Asymmetry arises due to spin-to-spin angular momentum conversion resulting in unequal output intensities of opposite circular polarizations for linearly polarized incident beam.

*Basic equations.* - The Maxwell equations for the electric field $\vec{E}\exp(-i\nu t)$ in a general inhomogeneous medium with dielectric constant $\varepsilon(x, y)$ reduce to:

$$\nabla^2\vec{E} + k^2 n^2 \vec{E} + \nabla\left(\vec{E}\cdot\ln n^2\right) = 0, \qquad (1)$$

where $k = 2\pi/\lambda$ is the wavenumber and $\varepsilon = n^2$ is the dielectric permittivity of the medium.

In the paraxial approximation, equation (1) can be reduced to the equivalent time-independent Schrodinger equation [13]. An analogous approach may be used to obtain a parabolic equation for the two-component vector field wavefunction [3]. Using the same method, the equation for a three-component wave equation can be derived:

$$\frac{i}{k}\frac{\partial\Psi}{\partial z} = \hat{H}\Psi, \qquad (2)$$

where

$$\Psi = n_0^{1/2}\exp(-ikn_0 z)\begin{bmatrix} e_x(r,\varphi) \\ e_y(r,\varphi) \\ e_z(r,\varphi) \end{bmatrix}, \quad \hat{H} = \hat{Z}^{-1}\left(\hat{H}_0 + \hat{H}_1\right) = \hat{H}_0 + \hat{H}_1 + \hat{H}_2,$$

$$\hat{H}_0 = \left[-\frac{1}{2k^2 n_0}\left(\frac{\partial^2}{\partial r^2} + \frac{1}{r}\frac{\partial}{\partial r} + \frac{1}{r^2}\frac{\partial^2}{\partial\varphi^2}\right) + \frac{1}{2n_0}\left(n_0^2 - n^2\right)\right]\hat{I}$$

is the unperturbed Hamiltonian corresponding to the first two terms in the equation (1),

$$\hat{H}_1 = -\frac{1}{2k^2 n_0}\begin{pmatrix} \frac{\partial}{\partial x}\cos\varphi\frac{\partial \ln n^2}{\partial r} & \frac{\partial}{\partial x}\sin\varphi\frac{\partial \ln n^2}{\partial r} & 0 \\ \frac{\partial}{\partial y}\cos\varphi\frac{\partial \ln n^2}{\partial r} & \frac{\partial}{\partial y}\sin\varphi\frac{\partial \ln n^2}{\partial r} & 0 \\ 0 & 0 & 0 \end{pmatrix}$$

and $\hat{H}_2 = \hat{Z}_1^{-1}\hat{H}_0$ are the perturbations corresponding to the third term in the equation (1),

$$\hat{Z}^{-1} = \begin{pmatrix} 1 & 0 & 0 \\ 0 & 1 & 0 \\ \frac{i}{2kn_0}\frac{\partial \ln n^2}{\partial x} & \frac{i}{2kn_0}\frac{\partial \ln n^2}{\partial y} & 1 \end{pmatrix} = \hat{I} + \hat{Z}_1^{-1}.$$

Consider a rotationally symmetric cylindrical waveguide with a parabolic distribution of the refractive index:

$$n^2(r) = n_0^2 - \omega^2 r^2 \ , \qquad (3)$$

where $n_0$ is the refractive index on the waveguide axis, $\omega$ is the gradient parameter, $r = \sqrt{x^2 + y^2}$ .

The Hamiltonian $\hat{H}$ may be rewritten in terms of annihilation and creation operators

$$\hat{A}_{1,2} = \frac{1}{\sqrt{2}}\left(\hat{a}_1 \pm i\hat{a}_2\right) \ , \qquad \hat{A}_{1,2}^{+} = \frac{1}{\sqrt{2}}\left(\hat{a}_1^{+} \mp i\hat{a}_2^{+}\right), \qquad \hat{a}_1 = \frac{1}{\sqrt{2}}\left(\sqrt{k\omega}\hat{x} + i\sqrt{\frac{k}{\omega}}\hat{p}_x\right),$$

$$\hat{a}_2 = \frac{1}{\sqrt{2}}\left(\sqrt{k\omega}\hat{y} + i\sqrt{\frac{k}{\omega}}\hat{p}_y\right), \ \hat{p}_x = -\frac{i}{k}\frac{\partial}{\partial x}, \hat{p}_y = -\frac{i}{k}\frac{\partial}{\partial y},$$

$$(x, y) = (r\cos\varphi, r\sin\varphi), \ \frac{\partial}{\partial x} = \cos\varphi\frac{\partial}{\partial r} - \frac{\sin\varphi}{r}\frac{\partial}{\partial\varphi}, \ \frac{\partial}{\partial y} = \sin\varphi\frac{\partial}{\partial r} + \frac{\cos\varphi}{r}\frac{\partial}{\partial\varphi}.$$

These operators satisfy the commutation relations: $\left[\hat{a}_i, \hat{a}_j^{+}\right] = \delta_{ij}$, $\left[\hat{A}_i, \hat{A}_j^{+}\right] = \delta_{ij}$.

Thus, we have

$$\hat{H}_0 = \frac{\omega}{kn_0}\left(\hat{A}_1^+\hat{A}_1 + A_2^+\hat{A}_2 + 1\right)\hat{I} \ ,$$

$$\hat{H}_1 = \eta\left[c_1\left(1+\frac{1}{2}\hat{\sigma}_z - \frac{1}{2}\hat{\sigma}_z^2\right) + c_2\left(\frac{1}{2}\hat{\sigma}_z + \frac{3}{2}\hat{\sigma}_z^2 - 1\right) + c_3\left(\hat{\sigma}_z\hat{\sigma}_+ - \hat{\sigma}_-\hat{\sigma}_z\right) + c_4\left(\sigma_z\sigma_+ + \sigma_-\sigma_z\right)\right], \quad (4)$$

$$\hat{H}_2 = \left[h\hat{\sigma}_-^2 + \frac{1}{2}s\left(\hat{\sigma}_- - \hat{\sigma}_-\hat{\sigma}_z - \hat{\sigma}_z\hat{\sigma}_-\right)\right]\hat{H}_0 \, .$$

Here

$$c_1 = \hat{1} + \hat{A}_1\hat{A}_2 - \hat{A}_1^+\hat{A}_2^+ \, , c_2 = \frac{1}{2}\left(\hat{A}_1^2 - \hat{A}_1^{+2} + \hat{A}_2^2 - \hat{A}_2^{+2}\right), \ c_3 = -ik\hat{L}_z \, , \ c_4 = -\frac{i}{2}\left(\hat{A}_1^2 + \hat{A}_1^{+2} - \hat{A}_2^2 - \hat{A}_2^{+2}\right),$$

$$h = -i\xi\left(\hat{A}_1 + \hat{A}_1^+ + \hat{A}_2 + \hat{A}_2^+\right), \qquad s = \xi\left(\hat{A}_1^+ - \hat{A}_1 + \hat{A}_2 - \hat{A}_2^+\right), \qquad \eta = \frac{\omega^2}{2k^2 n_0^3} \, , \qquad \xi = \frac{1}{2}\left(\frac{\omega}{k}\right)^{3/2}\frac{1}{n_0^3} \, ,$$

$$\hat{L}_z = -\frac{i}{k}\frac{\partial}{\partial\varphi} = \frac{1}{k}\left(\hat{A}_2^+\hat{A}_2 - \hat{A}_1^+\hat{A}_1\right), \ \hat{I} = \begin{pmatrix} 1 & 0 & 0 \\ 0 & 1 & 0 \\ 0 & 0 & 1 \end{pmatrix}$$ is the unit matrix and

$$\hat{\sigma}_x = \frac{1}{\sqrt{2}}\begin{pmatrix} 0 & 1 & 0 \\ 1 & 0 & 1 \\ 0 & 1 & 0 \end{pmatrix}, \qquad \hat{\sigma}_y = \frac{1}{\sqrt{2}}\begin{pmatrix} 0 & -i & 0 \\ i & 0 & -i \\ 0 & i & 0 \end{pmatrix}, \qquad \hat{\sigma}_z = \begin{pmatrix} 1 & 0 & 0 \\ 0 & 0 & 0 \\ 0 & 0 & -1 \end{pmatrix}, \qquad \hat{\sigma}_+ = \frac{1}{\sqrt{2}}\left(\hat{\sigma}_x + i\hat{\sigma}_y\right),$$

$$\hat{\sigma}_- = \frac{1}{\sqrt{2}}\left(\hat{\sigma}_x - i\hat{\sigma}_y\right).$$

Representation of the Hamiltonian by means of the operators will allow us to calculate the matrix elements analytically.

The solution of the equation (2) by the evolution operator $\hat{U} = \exp\left(-ik\hat{H}z\right)$ may be expressed as

$$\Psi(r,\varphi,z) = \hat{U}\Psi(r,\varphi,0) \, .$$

The solution of unperturbed equation is described by radially symmetric Gauss-Laguerre functions $\psi_{vl}(r,\varphi) = \left|v,l\right\rangle$, where $v = 2p + l$ is the principal quantum number, $p$ and $l$ are the radial and azimuthal indices, accordingly, and $l = v, v-2, v-4, \ldots 1$ or 0. The numbers $v$ and $l$ express the eigenvalues of the unperturbed Hamiltonian $\hat{H}_0 \left| v,l \right\rangle = \frac{\omega}{kn_0}(v+1)\left| v,l \right\rangle$, and eigenvalues $L = \frac{l}{k}$ of angular momentum operator $\hat{L}_z \left| v,l \right\rangle = \frac{l}{k}\left| v,l \right\rangle$.

*L - R asymmetry. -* The evolution of the coherency matrix is determined by the expression

$$\rho(z) = \hat{U}^+ \rho(0)\hat{U} \qquad (5)$$

or the equation

$$-\frac{i}{k}\frac{\partial \rho}{\partial z} = \hat{H}^+ \rho - \rho \hat{H} \ , \qquad (6)$$

where $\rho(0)$ is the operator of the coherency matrix at the initial plane $z = 0$ [3].

The total intensity of a light beam is given by

$$I(z) = Tr\rho(z) = Tr\left(\hat{U}^+ \rho(0)\hat{U}\right) \qquad (7)$$

Consider a linear polarized incident beam described by 3 x 3 coherency matrix (density matrix)

$$\rho(0) = \left| e_x \right\rangle\left\langle e_x \right| = \begin{pmatrix} \left| vl \right\rangle\left\langle vl \right| & 0 & \left| vl \right\rangle\left\langle e_z \right| \\ 0 & 0 & 0 \\ \left| e_z \right\rangle\left\langle vl \right| & 0 & \left| e_z \right\rangle\left\langle e_z \right| \end{pmatrix} = \qquad , \qquad (8)$$

$$\left| +1 \right\rangle \left| vl \right\rangle\left\langle vl \right| \left\langle +1 \right| + \left| +1 \right\rangle \left| vl \right\rangle\left\langle e_z \right| \left\langle -1 \right| + \left| -1 \right\rangle \left| e_z \right\rangle\left\langle vl \right| \left\langle +1 \right| + \left| -1 \right\rangle \left| e_z \right\rangle\left\langle e_z \right\langle -1 \right| \left. \right|$$

where $\left\langle e_x \right| = (\left\langle vl \right|, 0, \left\langle e_z \right|)$, $\left\langle +1 \right| = (1,0,0)$, $\left\langle 0 \right| = (0,1,0)$, $\left\langle -1 \right| = (0,0,1)$.

The longitudinal field component can be expressed through the transverse field components, i.e. $\left| e_z \right\rangle = \frac{i}{kn_0}\nabla_\perp \bar{e}_\perp$. Note that the polarization matrix (8) and the Hamiltonian (4) can be easily expressed also by means of spherical tensor operators $\hat{\tau}_{KQ}$. It was shown in [14] that vector and tensor operators completely describe the three-dimensional polarization. The mean values $t_{1Q} = \left\langle \hat{\tau}_{1Q} \right\rangle$ and $t_{2Q} = \left\langle \hat{\tau}_{2Q} \right\rangle$ describe the vector and tensor (rank 2) polarizations, accordingly.

Asymmetry can be defined as

$$\gamma = \frac{I^+ - I^-}{I^+ + I^-} = \frac{i\langle +1|\sum_{vl}\langle vl|U^+\rho(0)U|vl\rangle|0\rangle - i\langle 0|\sum_{vl}\langle vl|U^+\rho(0)U|vl\rangle|+1\rangle}{\langle +1|\sum_{vl}\langle vl|U^+\rho(0)U|vl\rangle|+1\rangle + \langle 0|\sum_{vl}\langle vl|U^+\rho(0)U|vl\rangle|0\rangle}, \tag{9}$$

where $I^+$ and $I^-$ are the intensities of right- and left-handed circularly polarized light, accordingly.

The following relations

$$\sigma_z|+1\rangle = |+1\rangle, \sigma_z|0\rangle = 0, \sigma_z|-1\rangle = -|-1\rangle; \sigma_+|+1\rangle = 0, \sigma_+|0\rangle = |+1\rangle, \sigma_+|-1\rangle = |0\rangle;$$

$$\sigma_-|+1\rangle = |0\rangle, \sigma_-|0\rangle = |-1\rangle, \sigma_-|-1\rangle = 0 \tag{10}$$

are used to calculate the matrix elements in (9).

Solving equation (6) within the accuracy of the small parameter $\eta$ and calculating the matrix elements in (9) we obtain the following expression for the asymmetry:

$$\gamma \cong -\frac{2l(v+1)}{n_0^4}\left(\frac{\omega}{k}\right)^2 \omega z \tag{11}$$

The gradient parameter of the waveguide defines the field radius of the fundamental mode (spot size) $w_0 = \sqrt{2/k\omega}$, so the asymmetry value can be expressed as

$$\gamma = \frac{I^+/I^- - 1}{I^+/I^- + 1} \approx -\frac{2l(v+1)}{8\pi^5 n_0^4} \cdot \frac{\lambda^5}{w_0^5} \cdot \frac{z}{w_0}. \tag{12}$$

This asymmetry is very small for conventional macroscopic optical fibers where $w_0 >> \lambda$, but becomes significant for fibers with diameter on the order of the light wavelength. The asymmetry increases with the increase of the orbital angular momentum (OAM) and the principal quantum number. Similar to the observation in nanoaperture [11], the sign of the asymmetry value depends on the sign of the beam vortex.

At $z = 0$ the intensities $I^+$ and $I^-$ are equal to each other, and the asymmetry is zero. The total intensity of a light beam is not changed at propagation, so the asymmetry occurs due to spin-to-spin angular momentum conversion.

*Discussion and conclusions.* - Let us estimate the value of asymmetry for the incident beam with the orbital angular momentum $l$ = 2 and the principal quantum number $v$ = 2. For the beam spot radius $w_0$ = 200 nm, wavelength $\lambda$ = 630 nm, and waveguide with $n_0$ = 1.5 and the length $z$ = 200 nm, we have asymmetry $\gamma \approx$ 0.3 and the transmission ratio $I^- / I^+ \approx 2$, which is easily observed.

Note that the OAM, in general, has either extrinsic or intrinsic parts [15]. The extrinsic OAM of a light beam is zero in our case, so only an intrinsic OAM is responsible for this effect. The term of the Hamiltonian $H_1$ containing the spin-orbit interaction does not contribute into the asymmetry, only the term $H_2$, which is the tensor (rank 2) interaction, causes such asymmetry. This indicates that the tensor forces are responsible for the asymmetry effect.

Thus the left-right asymmetry effect arises from second-rank polarization and not vector polarization, i.e. if the three-component field Maxwell equations are considered. There is no such asymmetry for conventional spin-orbit coupling of the type $\hat{L} \cdot \hat{s}$ following from two-component field equation. A quantum-mechanical analytical approach, developed here to solve the Maxwell equations, is shown to be efficient for investigation of 3D polarization evolution in a rotationally symmetric graded-index waveguide.

In conclusion, a novel fundamental effect of spin symmetry breaking via tensor interaction (spin-to-spin angular momentum transfer) is presented which occurs even in rotationally symmetric waveguides. Due to this effect the transformation of incident linear polarized light into particularly left- or right-handed circularly polarized light can be achieved. The sign of the helicity (left- or right-hand) of output light depends on the sign of the incident beam vortex phase. The necessary condition for this phenomenon is the existence of the longitudinal field component $e_z$ which becomes evident if the tightly collimated light beam is considered. The phenomenon could be exploited in polarization control applications, such as novel circular polarizers, polarization transformers and modulators.

References:

1. V.A. Fedotov, P.L. Mladyonov, S.L. Prosvirnin, A.V. Rogacheva, Y. Chen, and N. I. Zheludev, Phys. Rev. Lett. **97**, 167401 (2006).
2. E. Plum, X.X. Lin, V. Fedotov, Y. Chen, D.P. Tsai, N.I. Zheludev, Phys. Rev. Lett. **102**, 113902 (2009).
3. N.I. Petrov, J, Mod. Opt. **43**, 2239 (1996); Zh. Eksp. Teor. Fiz. **112**, 1985 (1997); Phys. Lett. A **234**, 239 (1997).
4. N.I. Petrov, Laser Physics **10**, 619 (2000).
5. K. Y. Bliokh, A. Niv, V. Kleiner, and E. Hasman, Nature Photonics **2**, 748 (2008).
6. Y. Gorodetsky, A. Niv, V. Kleiner, and E. Hasman, Phys. Rev. Lett. **101**, 043903 (2008).
7. K.Y. Bliokh, Y. Gorodetski, V. Kleiner, and E. Hasman, Phys. Rev. Lett. **101**, 030404 (2008).
8. V.S. Liberman, and B. Y. Zel'dovich, Phys. Rev. A **46**, 5199 (1992).
9. M. Onoda, S. Murakami, and N. Nagaosa, Phys. Rev. Lett. **93**, 083901 (2004).
10. A. Kavokin, G. Malpuech, and M. Glazov, Phys. Rev. Lett., **95** 136601 (2005).
11. Y. Gorodetski, N, Shitrit, I. Bretner, V. Kleiner, and E. Hasman, Nano Letters **9**, 3016 (2009).
12. L.T. Vuong, A.J.L. Adam, J.M. Brok, P.C.M. Planken, and H.P. Urbach, Phys. Rev. Lett. **104**, 083903 (2010).
13. V.A. Fock, and M.A. Leontovich, Zh. Eksp. Teor. Fiz. **16**, 557 (1946).
14. N.I. Petrov, Laser Physics **18**, 522 (2008).
15. A.T. O'Neil, I. MacVicar, L. Allen, and M.J. Padgett, Phys. Rev. Lett. **88**, 053601 (2002).